\documentclass[journal]{IEEEtran}
\usepackage{graphicx}
\usepackage{amsmath}
\usepackage{color}
\definecolor{orange}{rgb}{0,0,1}
\usepackage{multirow}
\usepackage{graphicx,float}
\usepackage{cite}
\ifCLASSINFOpdf
\else
  \DeclareGraphicsExtensions{.eps}
\fi
\usepackage[tight,footnotesize]{subfigure}
\begin{document}
%
\title{Performance Analysis of Strained Monolayer MoS$_{2}$ MOSFET}
%
%
%
\author{Amretashis Sengupta,~\IEEEmembership{Member,~IEEE,} Ram Krishna Ghosh, and Santanu Mahapatra,~\IEEEmembership{Senior~Member,~IEEE}  
\thanks{This work was supported by the Department of Science and Technology, Government of India, under grant no: SR/S3/EECE/0151/2012.}
\thanks{The authors are with the Nano-Scale Device Research Laboratory, Department of Electronic Systems Engineering, Indian Institute of Science, Bangalore-560012, India. (email: amretashis@dese.iisc.ernet.in, ramki.phys@gmail.com, santanu@cedt.iisc.ernet.in)}}
\maketitle
\begin{abstract}
We present a computational study on the impact of tensile/compressive uniaxial ($\varepsilon_{xx}$) and biaxial ($\varepsilon_{xx}=\varepsilon_{yy}$) strain on monolayer MoS$_{2}$ NMOS and PMOS FETs. The material properties like band structure, carrier effective mass and the multi-band Hamiltonian of the channel, are evaluated using the Density Functional Theory (DFT). Using these parameters, self-consistent Poisson-Schr\"{o}dinger solution under the Non-Equilibrium Green's Function (NEGF) formalism is carried out to simulate the MOS device characteristics. 1.75$\%$ uniaxial tensile strain is found to provide a minor (6$\%$) ON current improvement for the NMOSFET, whereas same amount of biaxial tensile strain is found to considerably improve the PMOSFET ON currents by 2-3 times. Compressive strain however degrades both NMOS and PMOS device performance. It is also observed that the improvement in PMOSFET can be attained only when the channel material becomes indirect-gap in nature. We further study the performance degradation in the quasi-ballistic long channel regime using a projected current method.
\end{abstract}
%
\IEEEpeerreviewmaketitle
\section{Introduction}
\IEEEPARstart{A}{among} the various classes of alternate channel materials under research, the 2-dimensional (2-D) materials having non-zero band gap in their sheet form like the Transition Metal Dichalcogenides (MX$_{2}$ : M=Mo, W; X=S, Se, Te) seem very promising for MOSFET applications. This is due to their better electrostatic integrity, optical transparency, mechanical flexibility and the geometrical compatibility with the standard planar CMOS technology. Among such MX$_{2}$ materials the performance of MoS$_{2}$ based MOS transistor and logic has been successfully demonstrated experimentally \cite{Radisavlejvic,Kis}. This has generated great interest in studying such `non-graphene' 2-D crystals for future MOSFET channel application \cite{Ganapathi,Alam,Guo}.\\
The main challenge in such 2-D MoS$_{2}$ FETs, so far has been to overcome the low carrier mobility of channel \cite{Radisavlejvic, Kis}. For Si CMOS, strain engineering has long been used to enhance carrier mobility and improve drive currents and other device parameters\cite{iwai}. Recent reports suggest that monolayer MoS$_{2}$ and other MX$_{2}$ also show alteration of material properties like band structure and carrier effective masses under the influence of strain\cite{Scalise, Yue, Penglu, Peelaers}. Also in their recent work Ghorbani-Asl et. al. \cite{Ghorbani} have shown the impact of strain on the conductance in MoS$_{2}$ sheets. Hence strain engineering in principle, could be used to improve the performance of MoS$_{2}$. In the present paper we investigate the impact of tensile and compressive uniaxial ($\varepsilon_{xx}$) and biaxial ($\varepsilon_{xx}=\varepsilon_{yy}$) strain on the performance of monolayer MoS$_{2}$ NMOS and PMOS devices. In our study, the material properties of 2-D (monolayer) MoS$_{2}$ , like band structure, carrier effective mass and the multi-band Hamiltonian of the channel, were evaluated using the density functional theory (DFT). Using these parameters, the MOS device output characteristics were simulated by solving the Poisson and the Schr\"{o}dinger equations self-consistently for the system, under the Non-Equilibrium Green's Function (NEGF) formalism. The device simulation results show only a minor performance enhancement for the NMOSFET under uniaxial tensile strain. On the other hand the PMOSFET performance is significantly improved by reducing the carrier effective mass by applying biaxial tensile strain.
\section{Methodology}
Fig. 1 shows the schematic device structure of the planar 2-D MoS$_{2}$ FET considered for our studies. We consider a monolayer MoS$_{2}$ as the channel material, with channel length($L_{Ch}$) of 10nm. As shown in Fig. 1, the tensile and the compressive strains are considered applied in the two perpendicular directions $x$ and $y$ in the plane of the 2-D sheet. For uniaxial strain only $\varepsilon_{xx}$ is applied whereas for the biaxial case strain is applied in both $x$ and $y$ direction with $\varepsilon_{xx}=\varepsilon_{yy}$. The 2-D channel is placed over an SiO$_{2}$/Si substrate. High-$\kappa$ HfO$_{2}$ of 2.5nm thickness is chosen as the gate dielectric. We consider highly doped ($10^{20}$ /$cm^{3}$) n$^{++}$ and p$^{++}$ regions as the source/drain for the NMOSFET and the PMOSFET respectively. Such doping concentrations allow for a very good alignment of the source/drain fermi levels with the conduction band/valence band for the monolayer MoS$_{2}$ NMOS and PMOS FETs\cite{Alam,Guo}. For our simulations $K \rightarrow \Gamma$ direction ($x$ direction in our Fig. 1) is taken as the transport direction.\\
The first step in our study is to evaluate the electronic properties of the channel material (i.e. strained and unstrained monolayer MoS$_{2}$ sheets). For this purpose we employ density functional theory(DFT) in QuantumWise ATK\cite{QW}. We use a 16$\times$16$\times$1 Monkhorst-Pack  k-grid\cite{Monkhorst,LDA} and employ the Localized Density Approximation (LDA)\cite {LDA} exchange correlation function with the Double-Zeta Polarized (DZP) basis\cite {Monkhorst}. We use Troullier-Martins type norm-conserving pseudopotential sets in ATK (NC-FHI[z=6] DZP for Sulfur and NC-FHI[z=6] DZP for Molybdenum). Relativistic corrections are included in the non-linear core.\cite{QW} Using DFT, we simulate the band structure and the electron and the hole effective masses of the monolayer MoS$_{2}$. The multi-band 41$\times$41 Hamiltonain matrix ($H$) and the non-orthogonal overlap matrix ($S$) are extracted from ATK at the valence band maxima (VB$_{max}$) and the conduction band minima (CB$_{min}$) of the band structure, for various strained and relaxed conditions. As with applied strain, the nature of the band gap of the monolayer changes from direct-gap to indirect-gap, we extract the Hamiltonians at the corresponding CB$_{min}$ for the NMOSFET, and at the corresponding VB$_{max}$ for the PMOSFET simulations.\\
\begin{figure}[H]
\begin{center}
\includegraphics[width=0.85\columnwidth]{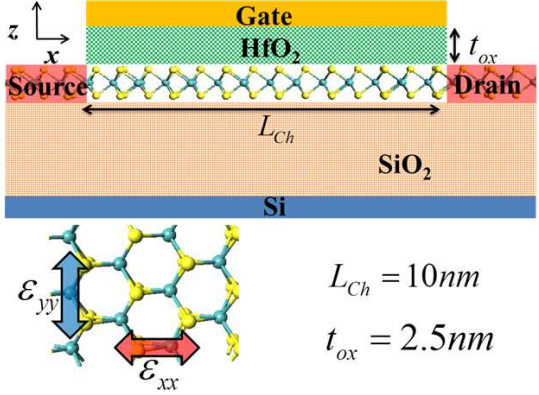}
\caption{Device schematic(not to scale) and diagram showing the applied uniaxial and biaxial tensile and compressive strains. We consider doped source and drains. $K \rightarrow \Gamma$ direction is taken as the transport direction.}
\end{center}
\label{fig_1}
\end{figure}
Thereafter, we proceed to solve the Poisson and Schr\"{o}dinger equations self-consistently for our MoS$_{2}$ FET. The self-consistent solutions are carried out under the Non-Equilibrium Green's Function (NEGF) formalism \cite {Dattabook,Dattapaper}. In our solver, we construct the Green's function from the knowledge of the $H$ and $S$ matrices and the energy eigenvalue matrix $E$ of the system along with the self-energy matrices $\Sigma_{S}$ and $\Sigma_{D}$ of the source and drain contacts respectively. The Green's function is then evaluated as \cite{Dattabook}
\begin{equation}
G(E) = [ES-H-\Sigma_{S}-\Sigma_{D}]^{-1}
\label{equn_1}
\end{equation}
From (1) parameters like the broadening matrices $\wp_{S}$ and $\wp_{D}$ and the spectral densities $A_{S}$ and $A_{D}$ are evaluated using the relations
\begin{equation}
\wp_{S,D}=i[\Sigma_{S,D}-\Sigma^{+}_{S,D}]
\label{equn_2}
\end{equation}
\begin{equation}
A_{S,D}=G\wp_{S,D}G^{+}
\label{equn_3}
\end{equation}
The density matrix [$\Re$] used to solve the Poisson equation is given by
\begin{equation}
[\Re] = \int_{-\infty}^\infty{\frac{dE}{2\pi}[A(E_{k,x})]f_0(E_{k,x}-\eta)}
\label{equn_4}
\end{equation}
where $A(E_{k,x})$ is the spectral density matrix,$E_{k,x}$ the energy of the conducting level,and $\eta$ being the chemical potential of the contacts. $f_{0}(.)$ is the Fermi function. The transmission matrix $\Im(E)$ is calculated as
\begin{equation}
\Im(E)=trace[A_{S}\wp_{D}]=trace[A_{D}\wp_{S}]
\label{equn_5}
\end{equation}
Thus giving out the ballistic drain current $I_{D,Bal}$ as \cite{Ganapathi}
\begin{multline}
I_{D,Bal}={\frac{q}{\hbar^{2}}}\sqrt{\frac{m_{t}\varphi_{Th}}{2{\pi}^{3}}}
\int_{-\infty}^\infty[F_{-1/2}(\frac{\eta_{S}-E_{k,x}}{\varphi_{Th}})\\-F_{-1/2}(\frac{\eta_{D}-E_{k,x}}{\varphi_{Th}})]\Im(E_{k,x})dE
\label{equn_6}
\end{multline}
$m_{t}$ being the carrier effective mass in the transverse direction, $\varphi_{Th}$ is the thermal energy, $E_{k,x}$ the energy of the conducting level, $F_{-1/2}$ is the Fermi integral of order $-1/2$. $\eta_{S}$ and $\eta_{D}$ are the chemical potentials of the source and drain respectively. It is notable that the current calculated in (6) is purely ballistic in nature, which holds well for channels of short dimensions upto few tens of nanometers. However for longer channel lengths the transmission encounters scattering, and becomes quasi-ballistic in nature. For considering these effects, we use a projection factor $\Theta$ to evaluate our MOSFET drain current as \cite {Ganapathi,Alam}
\begin{equation}
I_{D}=\Theta \times I_{D,Bal}
\label{equn_7}
\end{equation}
the value of $\Theta$ is determined as
\begin{equation}
\Theta=\frac{\lambda_{max}}{L_{Ch}+\lambda_{max}}
\label{equn_8}
\end{equation}
Where $L_{Ch}$ is the channel length and $\lambda_{max}$ is the mean free path calculated as \cite {Ganapathi,Alam}
\begin{equation}
\lambda_{max}=\frac{{(2\varphi_{Th})}^{3/2}}{q\mu}\frac{F_{0}(\eta_{S}-E_{C})}{F_{-1/2}(\eta_{S}-E_{C})}
\label{equn_9}
\end{equation}
Here, $E_{C}$ is the top of the conduction band energy in the channel, which is evaluated from the maxima of the self-consistent potential $\Phi_{SC}$ in the channel, $F_{-1/2}$ is the 1-D Fermi integral of order $-1/2$, $\mu$ is the carrier mobility. It is worth noting that for short channel lengths, $\lambda_{max}$ $\gg$ $L_{Ch}$, and therefore $\Theta$ $\rightarrow$ 1, which is the purely ballistic case.
\section{Results and Discussions}
\subsection{Materials study with DFT}
In Fig. 2, we have shown the impact of various uniaxial and biaxial tensile and compressive strains on the band gap of the monolayer MoS$_{2}$. In the relaxed case the band structure is direct in nature, with the VB$_{max}$ and CB$_{min}$ both at the K point of the Brillouin Zone(BZ). However as we apply strain to the system the band structure changes and the MoS$_{2}$ undergoes transition from direct-gap to an indirect gap material. It is seen that for uniaxial strain the material becomes indirect gap at tensile strain of +1.25$\%$ whereas for biaxial strain it becomes indirect for strains above +0.75$\%$. For uniaxial compressive strain the band structure remains direct at the K point for strains upto -1.25$\%$ but for biaxial compressive strain of -1.25$\%$ the material becomes indirect gap in nature. We have not shown compressive strains further than -1.25$\%$ as this increases the carrier effective mass (not shown here) and therefore is degenerative to device performance, as we shall see in the subsequent section.\\
\begin{figure}[H]
\begin{center}
\includegraphics[width=1.0\columnwidth]{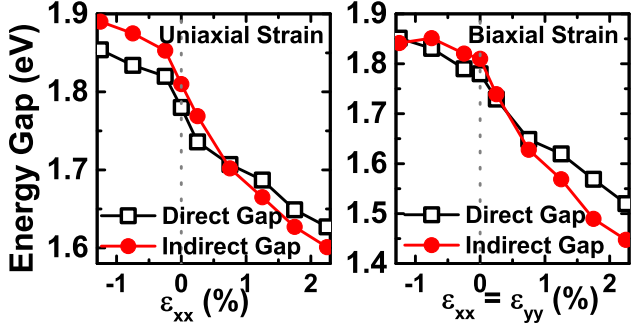}
\caption{Effect of uniaxial($\varepsilon_{xx}$) and biaxial($\varepsilon_{xx}=\varepsilon_{yy}$) tensile ($+$) and compressive ($-$) strain on the band-gap of the monolayer MoS$_{2}$.}
\end{center}
\label{fig_2}
\end{figure}
The direct band-gap is always measured at the K point of the BZ, for relaxed MoS$_{2}$ sheet it was found to be 1.78eV, which is consistent with other DFT results\cite{Peelaers}. In the relaxed condition the monolayer MoS$_{2}$ shows a slightly higher indirect gap of 1.82eV between the CB$_{min}$ at K point and the VB$_{max}$ at $\Gamma$ point. In the different strained condition the indirect gap however is measured between the the different sets of VB$_{max}$ and CB$_{min}$ as the band structure changes. In Fig. 3 we see that for tensile uniaxial and biaxial strain, the VB$_{max}$ is at the $\Gamma$ point while the CB$_{min}$ remains at the K point. However for the uniaxial and the biaxial compressive strain, the VB$_{max}$ remains fixed at the K point but the CB$_{min}$ tends to shift to a point in between the K point and the $\Gamma$ point (i.e. in the $\Lambda$ direction of the hexagonal BZ) which we shall designate as $\Lambda_{min}$ hereafter. These band structure results are in good agreement with DFT results published by other groups\cite{Peelaers,Penglu,Kuc}.\\
\begin{table}[!htbp]
\renewcommand{\arraystretch}{1.0}
\caption{Electronic band properties under uniaxial strain}
\label{table_uniax}
\centering
\begin{tabular}{l c c c c c r}
\hline
\bfseries $\varepsilon$($\%$) & \bfseries VB$_{max}$ & \bfseries CB$_{min}$ & \bfseries $m_{e}$ & \bfseries $m_{e,t}$ & \bfseries $m_{h}$ & \bfseries $m_{h,t}$\\
\hline\hline
-1.25 & $K$ & $K$ & $K(\Lambda)$ & $K(T)$ & $K(\Lambda)$ & $K(T)$\\
-0.75 & $K$ & $K$ & $K(\Lambda)$ & $K(T)$ & $K(\Lambda)$ & $K(T)$\\
-0.25 & $K$ & $K$ & $K(\Lambda)$ & $K(T)$ & $K(\Lambda)$ & $K(T)$\\
0 & $K$ & $K$ & $K(\Lambda)$ & $K(T)$ & $K(\Lambda)$ & $K(T)$\\
+0.25 & $K$ & $K$ & $K(\Lambda)$ & $K(T)$ & $K(\Lambda)$ & $K(T)$\\
+0.75 & $\Gamma$ & $K$ & $K(\Lambda)$ & $K(T)$ & $\Gamma(\Lambda)$ & $\Gamma(\Sigma)$\\
+1.25 & $\Gamma$ & $K$ & $K(\Lambda)$ & $K(T)$ & $\Gamma(\Lambda)$ & $\Gamma(\Sigma)$\\
+1.75 & $\Gamma$ & $K$ & $K(\Lambda)$ & $K(T)$ & $\Gamma(\Lambda)$ & $\Gamma(\Sigma)$\\
+2.25 & $\Gamma$ & $K$ & $K(\Lambda)$ & $K(T)$ & $\Gamma(\Lambda)$ & $\Gamma(\Sigma)$\\
\hline
\end{tabular}
\end{table}
\begin{table}[!htbp]
\renewcommand{\arraystretch}{1.0}
\caption{Electronic band properties under biaxial strain}
\label{table_biax}
\centering
\begin{tabular}{l c c c c c r}
\hline
\bfseries $\varepsilon$($\%$) & \bfseries VB$_{max}$ & \bfseries CB$_{min}$ & \bfseries $m_{e}$ & \bfseries $m_{e,t}$ & \bfseries $m_{h}$ & \bfseries $m_{h,t}$\\
\hline\hline
-1.25 & $K$ & $\Lambda_{min}$ & $\Lambda_{min}(\Lambda)$ & $\Lambda_{min}(T)$ & $K(\Lambda)$ & $K(T)$\\
-0.75 & $K$ & $K$ & $K(\Lambda)$ & $K(T)$ & $K(\Lambda)$ & $K(T)$\\
-0.25 & $K$ & $K$ & $K(\Lambda)$ & $K(T)$ & $K(\Lambda)$ & $K(T)$\\
0 & $K$ & $K$ & $K(\Lambda)$ & $K(T)$ & $K(\Lambda)$ & $K(T)$\\
+0.25 & $K$ & $K$ & $K(\Lambda)$ & $K(T)$ & $K(\Lambda)$ & $K(T)$\\
+0.75 & $\Gamma$ & $K$ & $K(\Lambda)$ & $K(T)$ & $\Gamma(\Lambda)$ & $\Gamma(\Sigma)$\\
+1.25 & $\Gamma$ & $K$ & $K(\Lambda)$ & $K(T)$ & $\Gamma(\Lambda)$ & $\Gamma(\Sigma)$\\
+1.75 & $\Gamma$ & $K$ & $K(\Lambda)$ & $K(T)$ & $\Gamma(\Lambda)$ & $\Gamma(\Sigma)$\\
+2.25 & $\Gamma$ & $K$ & $K(\Lambda)$ & $K(T)$ & $\Gamma(\Lambda)$ & $\Gamma(\Sigma)$\\
\hline
\end{tabular}
\end{table}
\begin{figure}[H]
\begin{center}
\includegraphics[width=1.0\columnwidth]{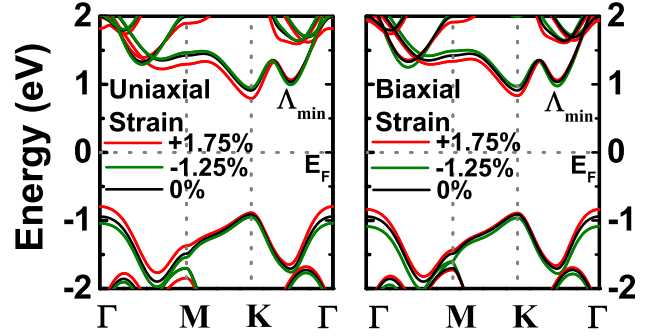}
\caption{The band structure of the MoS$_{2}$ sheet under different strain conditions.}
\end{center}
\label{fig_3}
\end{figure}
Fig. 4 shows the variation of the carrier effective masses with applied strain at the different symmetry points ($\Gamma$, $K$ and $\Lambda_{min}$) of the BZ in different crystallographic directions. The $K \rightarrow \Gamma$ direction is referred to as the $\Lambda$ direction, the $K \rightarrow M$  direction as $T$ and $\Gamma \rightarrow M$ direction as $\Sigma$. Thus the legend $K(\Lambda)$ in Fig. 4, represents the carrier effective mass at the $K$ point of the BZ in the $K \rightarrow \Gamma$ direction and $\Lambda_{min}(\Lambda)$ represents the carrier effective mass at the $\Lambda_{min}$ point in the same direction, and so on. We can see that with the application of tensile strain, there is a slight reduction in the $K(\Lambda)$ and the $K(T)$ electron effective masses for both uniaxial and biaxial conditions. However with uniaxial compressive strain, the electron effective masses increase. For biaxial compressive strain there is an increment in $K(\Lambda)$ and the $K(T)$ electron effective masses but a decrease in the $\Lambda_{min}(\Lambda)$ and $\Lambda_{min}(T)$ electron masses. As for the hole effective masses, there is not much change in the $K(\Lambda)$ and the $K(T)$ hole masses for uniaxial or biaxial tensile and compressive strain. However, for the hole effective mass in the $\Gamma(\Lambda)$ and the $\Gamma(\Sigma)$ there exists a significant change for biaxial strain. The values of electron and hole masses in the relaxed MoS$_{2}$ for $K(\Lambda)$ are 0.4750$m_{0}$ and 0.5978$m_{0}$ respectively. For $K(T)$ these values are 0.4741$m_{0}$ and 0.5968$m_{0}$ respectively. These results are consistent with other ab-inito studies\cite{Scalise,Penglu,Peelaers}. With application of +2.25$\%$ biaxial strain, the hole effective mass could be brought down by 41$\%$ from its relaxed value. While for the electron a +2.25$\%$ uniaxial strain reduces the effective mass only by 3$\%$. These values and the nature of the variation of electron and hole effective masses with uniaxial and biaxial strain on are consistent with the ab-initio results published by others\cite{Peelaers}.\\
\begin{figure*}[!htbp]
\begin{center}
\subfigure[]{\includegraphics[width=0.75\columnwidth]{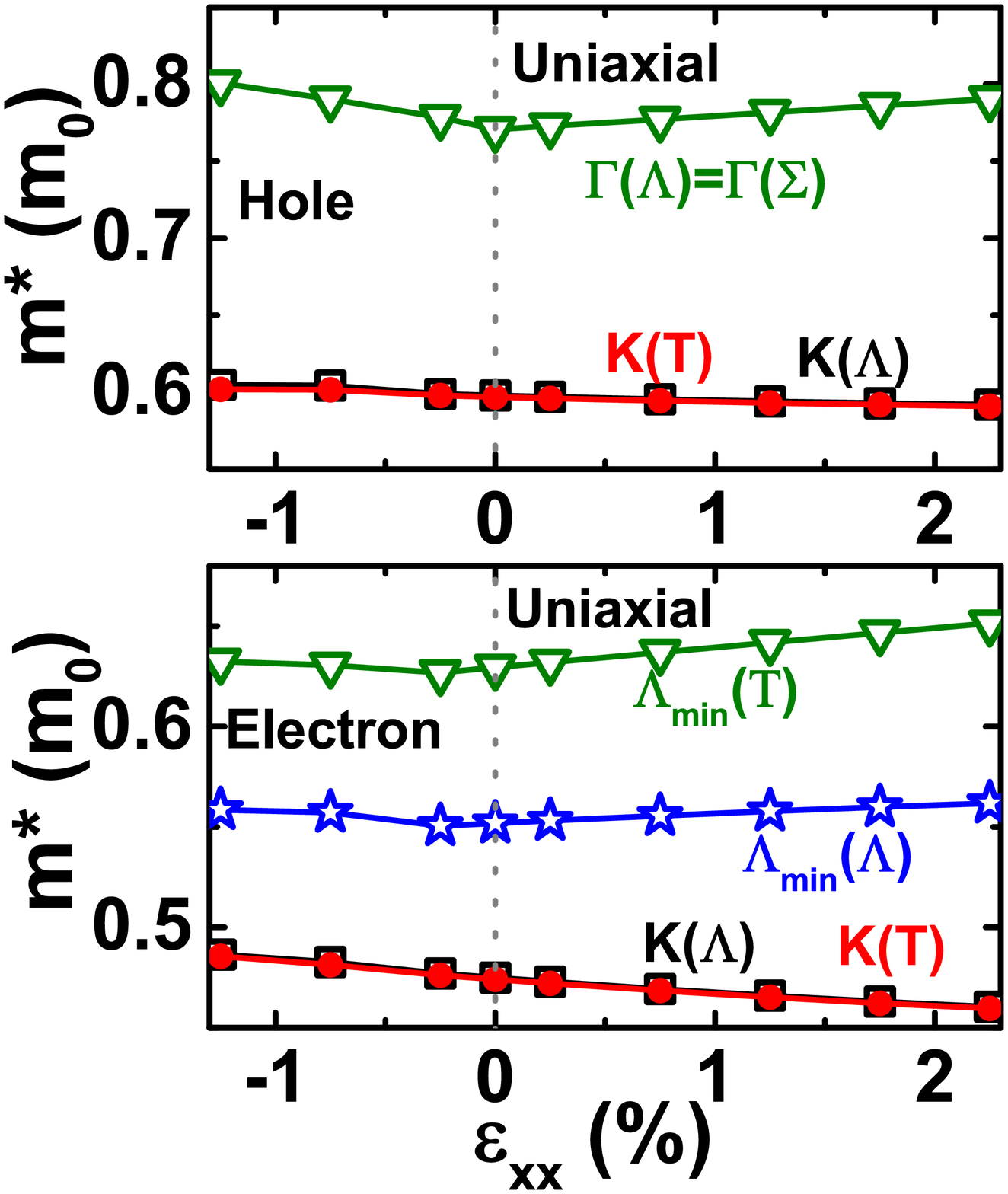}\label{a}}
\subfigure[]{\includegraphics[width=0.75\columnwidth]{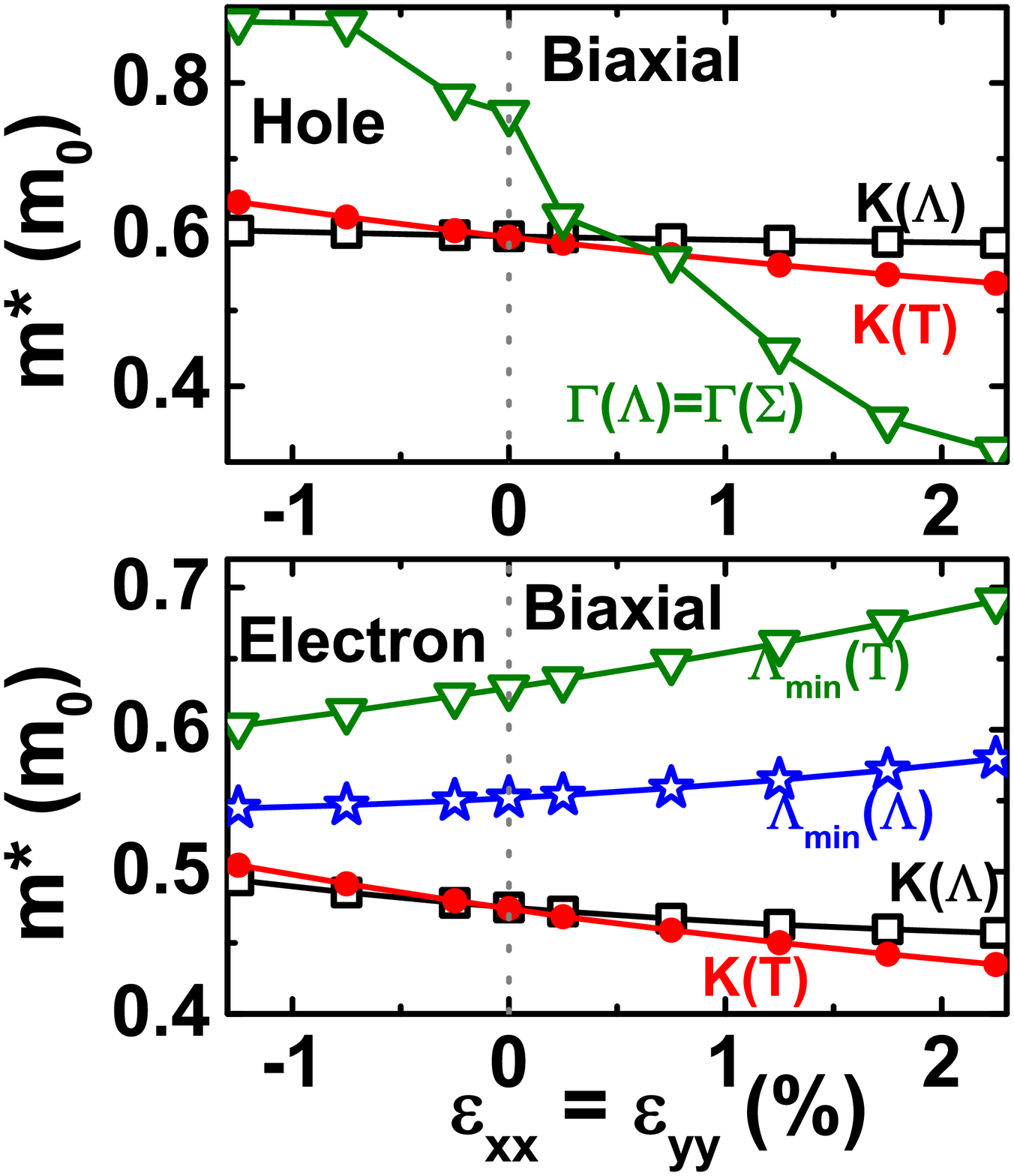}\label{b}}
\caption{The electron and the hole effective masses in the monolayer MoS$_{2}$ channel for (a) uniaxial and (b)biaxial strain.}
\label{fig_4}
\end{center}
\end{figure*}
\begin{figure*}[!htbp]
\begin{center}
\subfigure[]{\includegraphics[width=0.75\columnwidth]{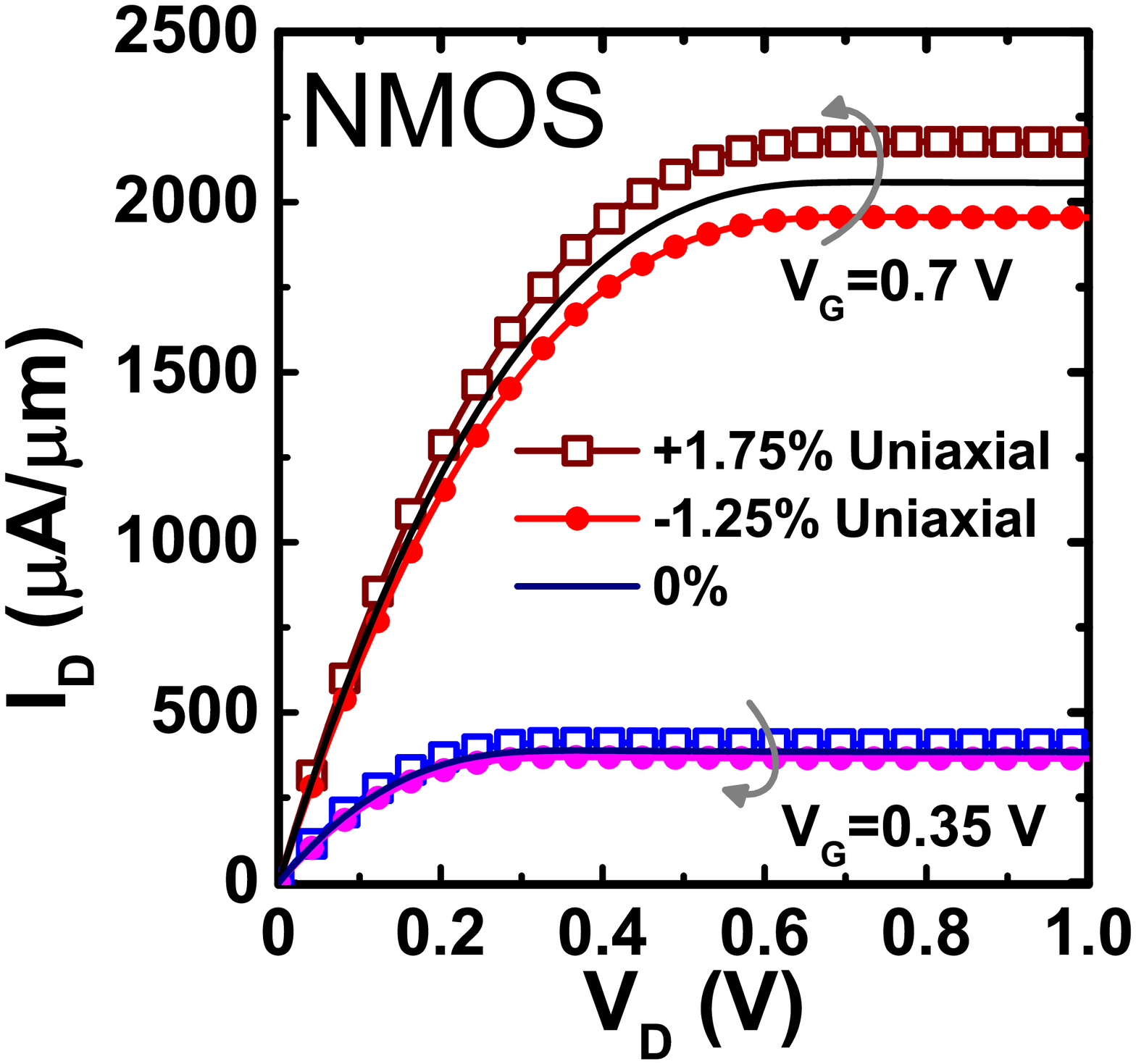}\label{a}}
\subfigure[]{\includegraphics[width=0.75\columnwidth]{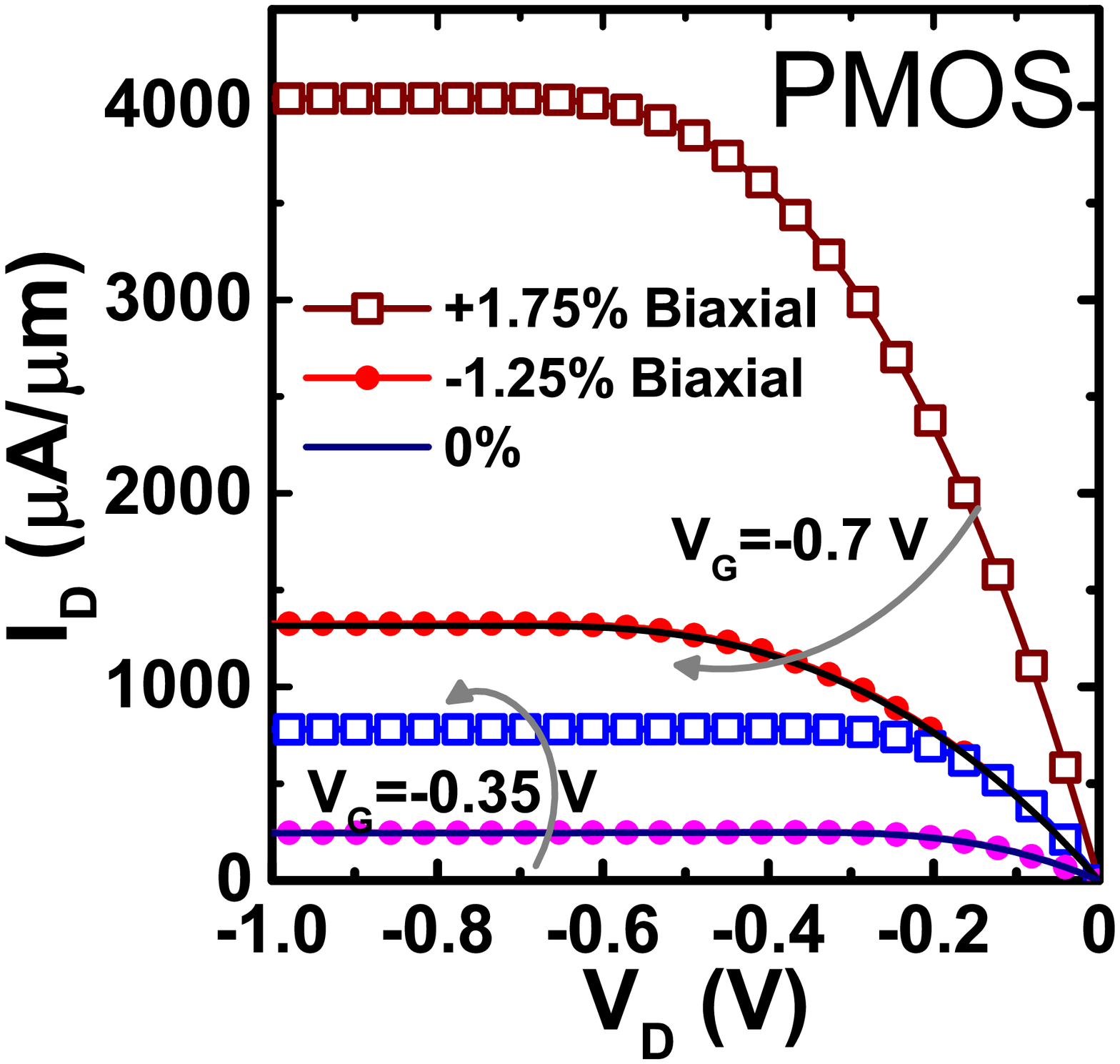}\label{b}}
\subfigure[]{\includegraphics[width=0.75\columnwidth]{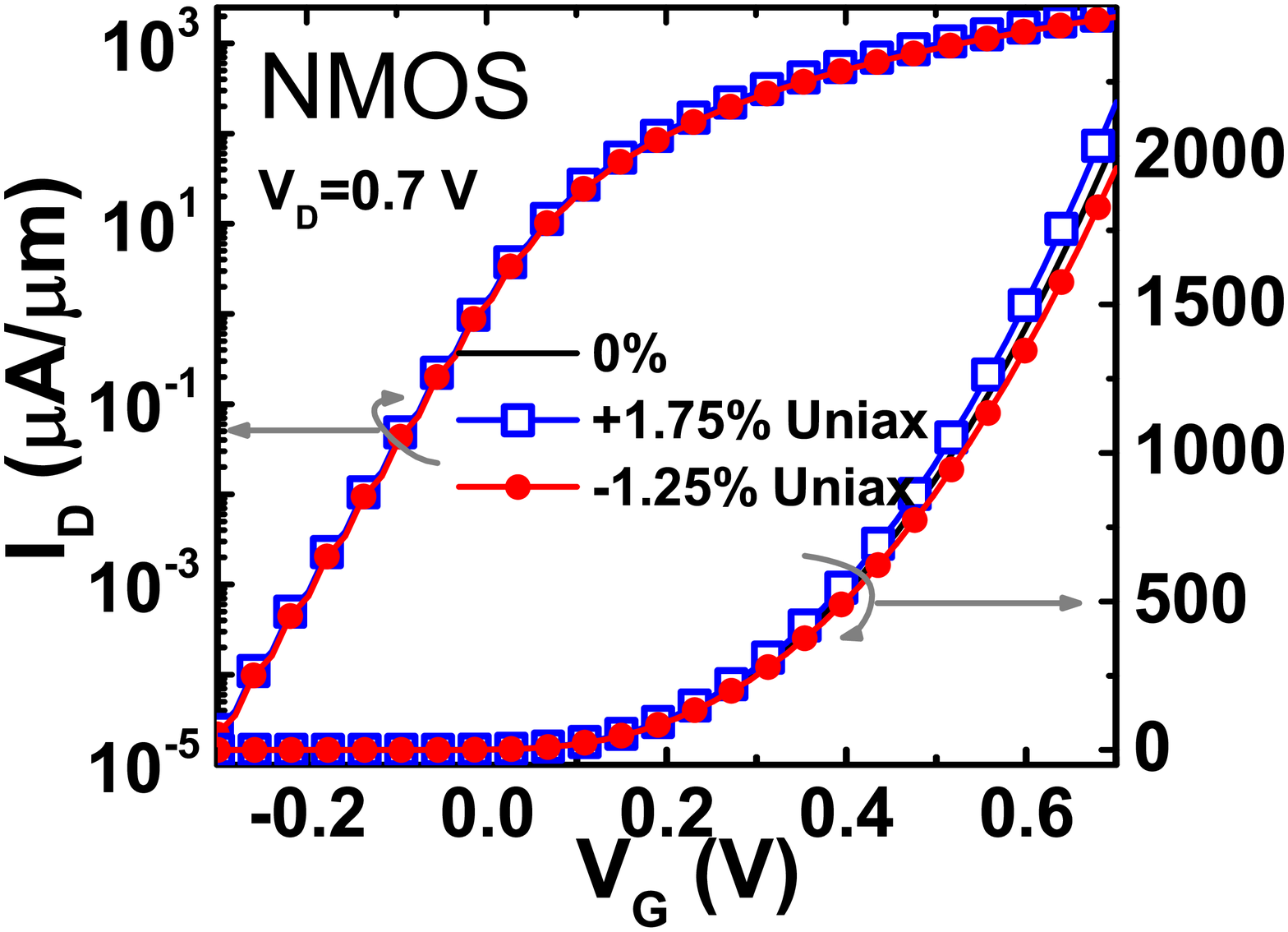}\label{c}}
\subfigure[]{\includegraphics[width=0.75\columnwidth]{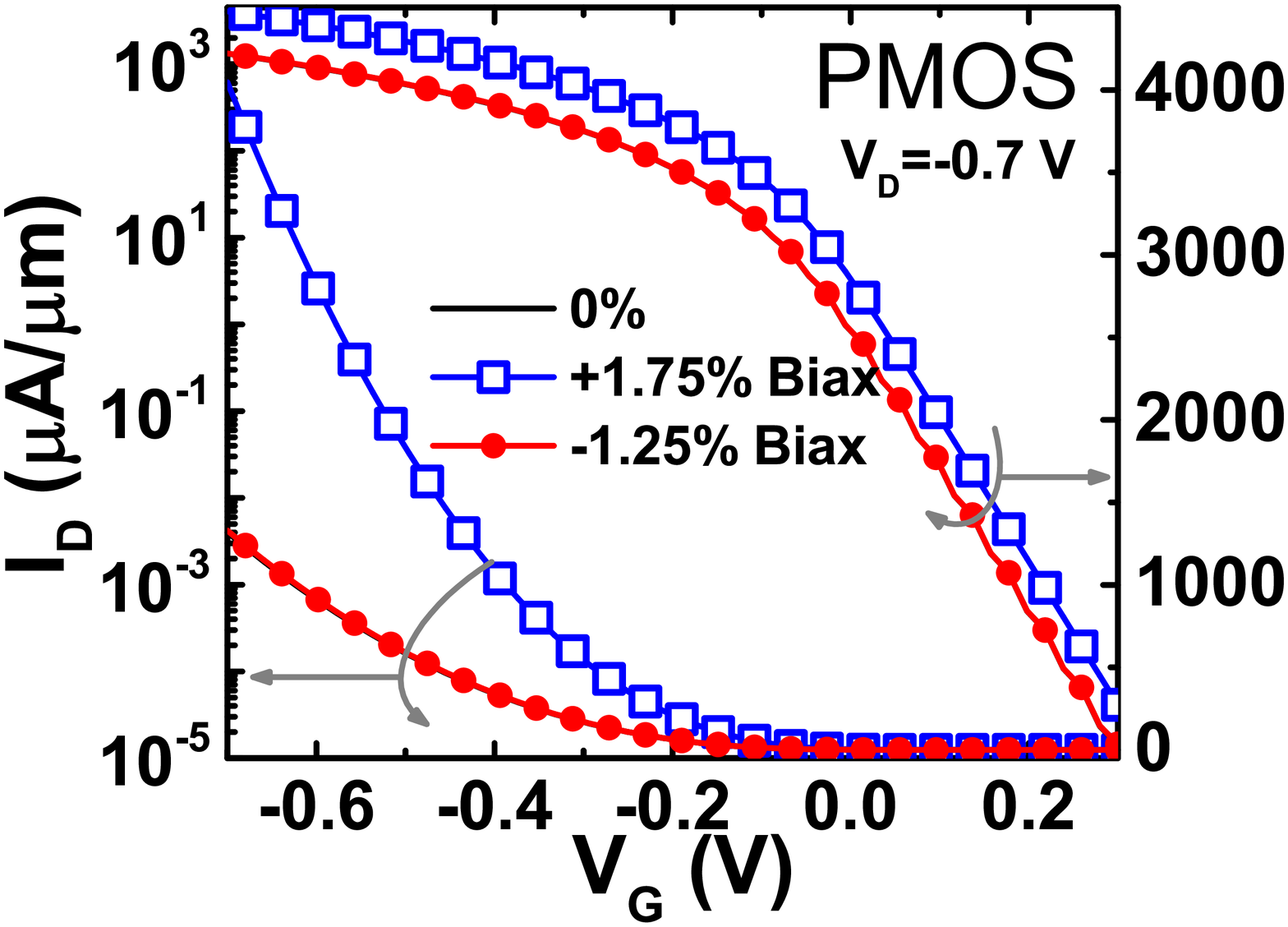}\label{d}}
\caption{The (a,b) $I_{D}$-$V_{D}$  and the (c,d) $I_{D}$-$V_{G}$ characteristics of the NMOS and PMOS devices, with varying strain conditions.}
\end{center}
\label{fig_5}
\end{figure*}
\subsection{Device simulation}
Tables I and II, shows the location of the different VB$_{max}$ and CB$_{min}$ under varying uniaxial and biaxial strain along with the corresponding carrier masses that need to be considered for device simulation. Here, $m_{e}$ and $m_{e,t}$ represents the electron masses in the transport and the transverse direction respectively, while $m_{h}$ and $m_{h,t}$ represent the hole masses for the same directions. The $H$ and $S$ matrices are extracted at those particular VB$_{max}$ and CB$_{min}$, for PMOS and NMOS simulation respectively.\\
Since for the performance enhancement of MoS$_{2}$ FET devices, the lowering of the carrier effective mass is essential\cite{Guo}, hence for the simulations we focus on the strains which decrease the carrier effective masses in our MOSFET devices. For the NMOSFET, we consider the carrier masses under the uniaxial strain condition and for the PMOSFET we consider the biaxial strained condition.\\
The static dielectric constant $Re[\epsilon(\omega=0)]$ of MoS$_{2}$ is evaluated from the optical spectra in ATK to be 3.92. This value is not affected by the applied strain (not shown here).\\
The I$_{D}$-V$_{D}$ output characteristics (Fig. 5) of the NMOSFET and PMOSFET devices shows the variation of drain current for varying applied strain and gate voltages. For comparison of the tensile and the compressive strains, we have shown the devices under three conditions which are relaxed (0$\%$ strain), +1.75$\%$ strained and -1.25$\%$ strained channels. As mentioned earlier the nature of applied strain is uniaxial for the NMOSFET and biaxial for the PMOSFET respectively.\\
The drive current value for the relaxed MoS$_{2}$ NMOSFET is about 2058 $\mu$A/$\mu$m and that for the PMOSFET is 1545 $\mu$A/$\mu$m, which is quite sufficient for the ITRS requirements for the 15nm and lesser high-performance (HP) logic technology node \cite {itr}. We see that with application of +1.75$\%$ uniaxial strain the ON current for NMOSFET could be increased to 2178 $\mu$A/$\mu$m, which is a 5.83$\%$ improvement over the relaxed value. However for the -1.25$\%$ uniaxial strain, the NMOS ON current decreases by about 4$\%$. For the PMOSFET, a very significant improvement is observed upon application of +1.75$\%$ biaxial strain. For this strain, the ON current becomes 4041 $\mu$A/$\mu$m, which is a two and a half fold increase over the relaxed ON current. In case of PMOS a slight degradation of ON current is observed for -1.25$\%$ biaxial strain. For +1.75$\%$ biaxial strain, the performance of the monolayer MoS$_{2}$ PMOSFET can be greatly improved. In comparison to the PMOSFET, the improvement in the NMOSFET for an equal amount of uniaxial strain is just 6$\%$.\\
From our simulations it is also observed that for both the NMOSFET and the PMOSFET, applied strain does not impact the subthreshold slope (SS) significantly. Both strained and relaxed NMOSFET and PMOSFET, show good immunity to short channel effects, with DIBL within the range of 12-15 mV/V. The SS is calculated to be 60-62.5 mV/decade. The ON/OFF ratio is determined to be $~10^{8}$ considering V$_{dd}$=0.7 V. These values are better than those for the FD SOI and the MG MOSFET, for the 15nm high-performance(HP) logic node as recommended by ITRS \cite {itr}.\\
\begin{figure}[!ht]
\begin{center}
\subfigure[]{\includegraphics[width=1.7in]{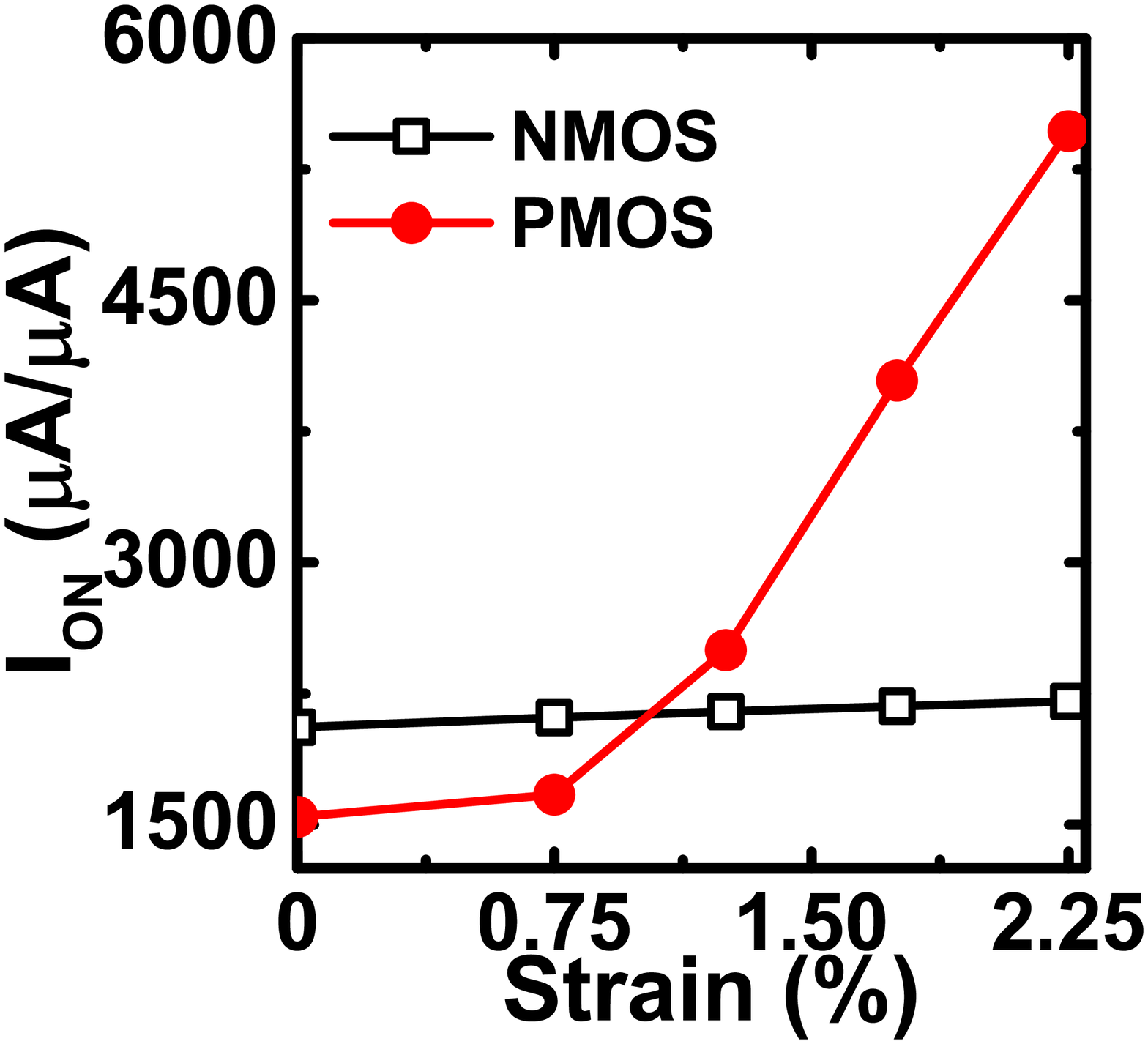}\label{a}}
\subfigure[]{\includegraphics[width=1.69in]{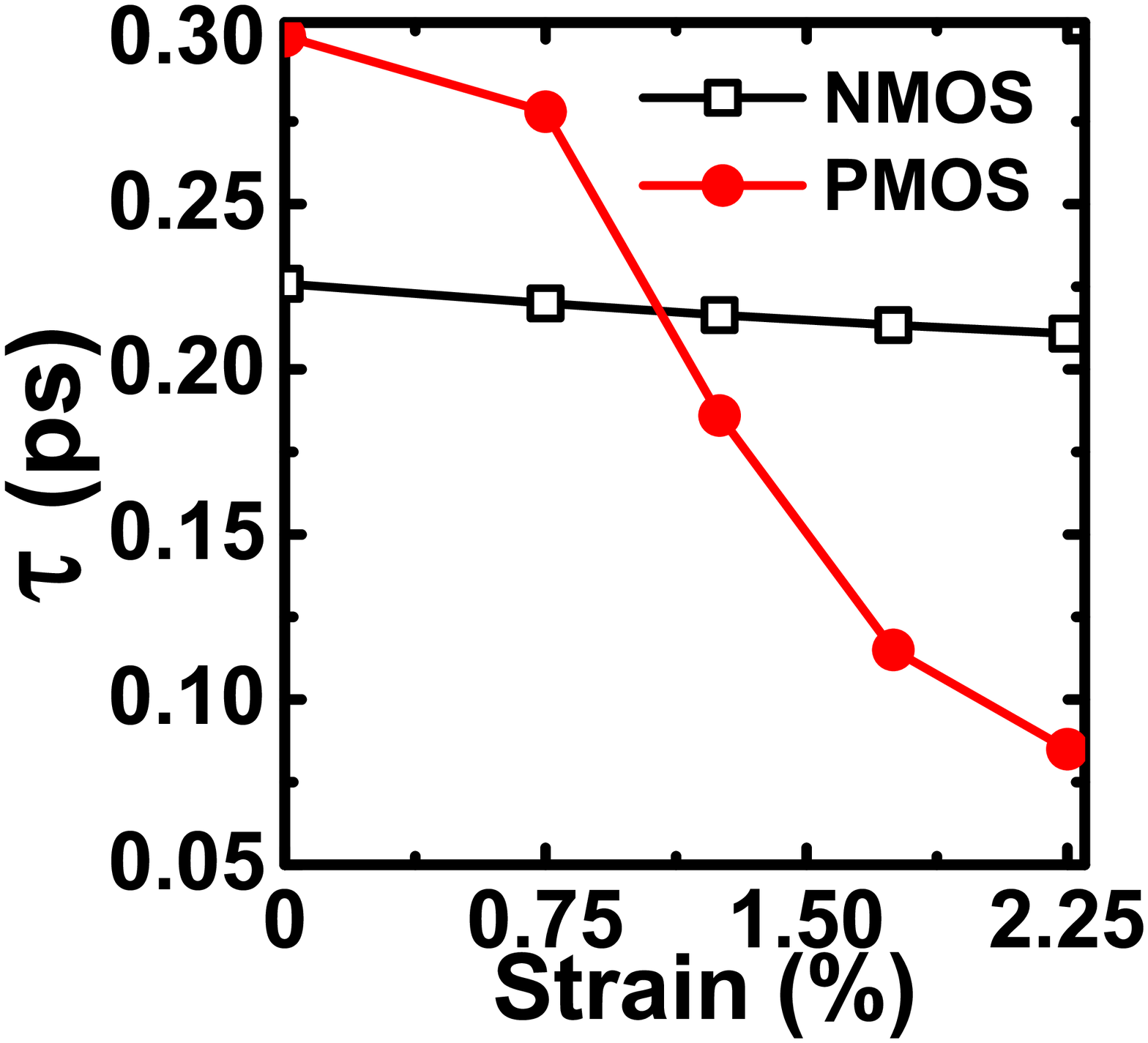}\label{b}}
\caption{Impact of strain on the (a) ON current and (b) the intrinsic delay time of the MoS$_{2}$ NMOS and PMOS FETs. Uniaxial strain is considered for NMOSFET and biaxial strain for PMOSFET.}
\end{center}
\label{fig_6}
\end{figure}
\begin{figure}[!hb]
\begin{center}
\includegraphics[width=1.0\columnwidth]{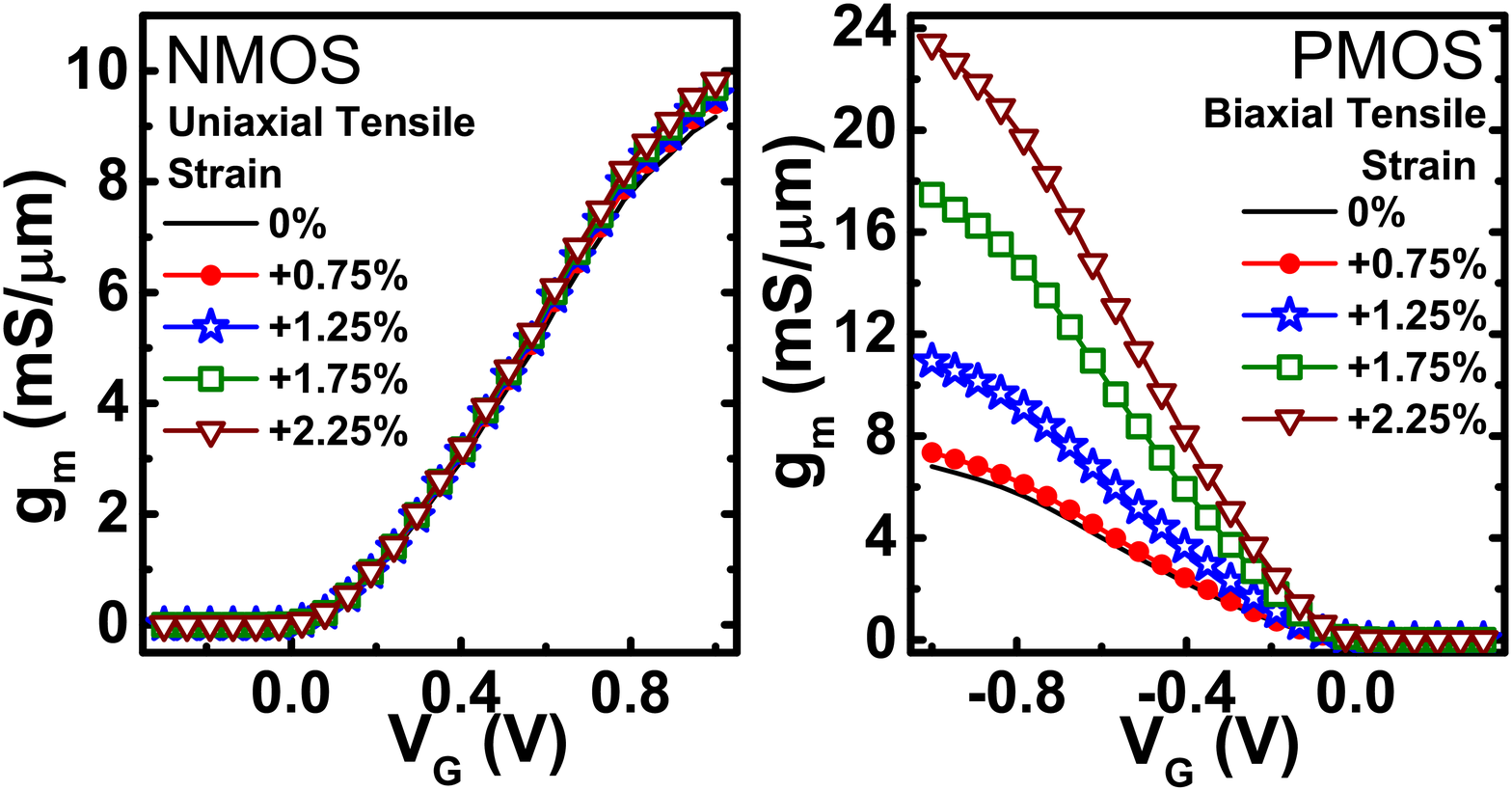}
\caption{Impact of strain on the transconductance ($g_{m}$) vs V$_{G}$ plot of the MoS$_{2}$ NMOS and PMOS FETs. }
\end{center}
\label{fig_7}
\end{figure}
The ON current and the intrinsic delay time improvement with applied tensile strain for the MoS$_{2}$ FETs are shown in Fig. 6. With increasing uniaxial tensile strain for the NMOS device, and biaxial tensile strain for the PMOS, significant improvement is observed in the ON currents and the delay time ($\tau$) of these devices. For uniaxial strains of +2.25$\%$ an increment of 7.2$\%$ can be brought about for NMOS, whereas for the PMOS, biaxial strain of the same magnitude can increase the ON currents by 3.6 times its relaxed value. For the same applied strain, the corresponding reduction in $\tau$ is about 18$\%$ for the NMOSFET and almost 80$\%$ for the PMOSFET.\\
In Fig. 7, we have shown the transconductance ($g_{m}$) versus gate voltage for the MOSFETs under consideration. For the simulation $V_{D}$ is set at 0.7 V. The value of $g_{m}$ at the ON condition ($V_{D}=V_{G}=0.7 V$), for the NMOSFET and the PMOSFET in the relaxed condition are 7.2 mS/$\mu$m and 7.5 mS/$\mu$m respectively. For the strained condition, the $g_{m}$ slightly increases upto 8 mS/$\mu$m for the NMOSFET. For the PMOSFET, the increment in $g_{m}$ is much more prominent and in the ON condition, the value of $g_{m}$ for +2.25$\%$ biaxially strained PMOSFET reaches to about 20.5 mS/$\mu$m.\\
\begin{figure}[!ht]
\begin{center}
\subfigure[]{\includegraphics[width=1.7in]{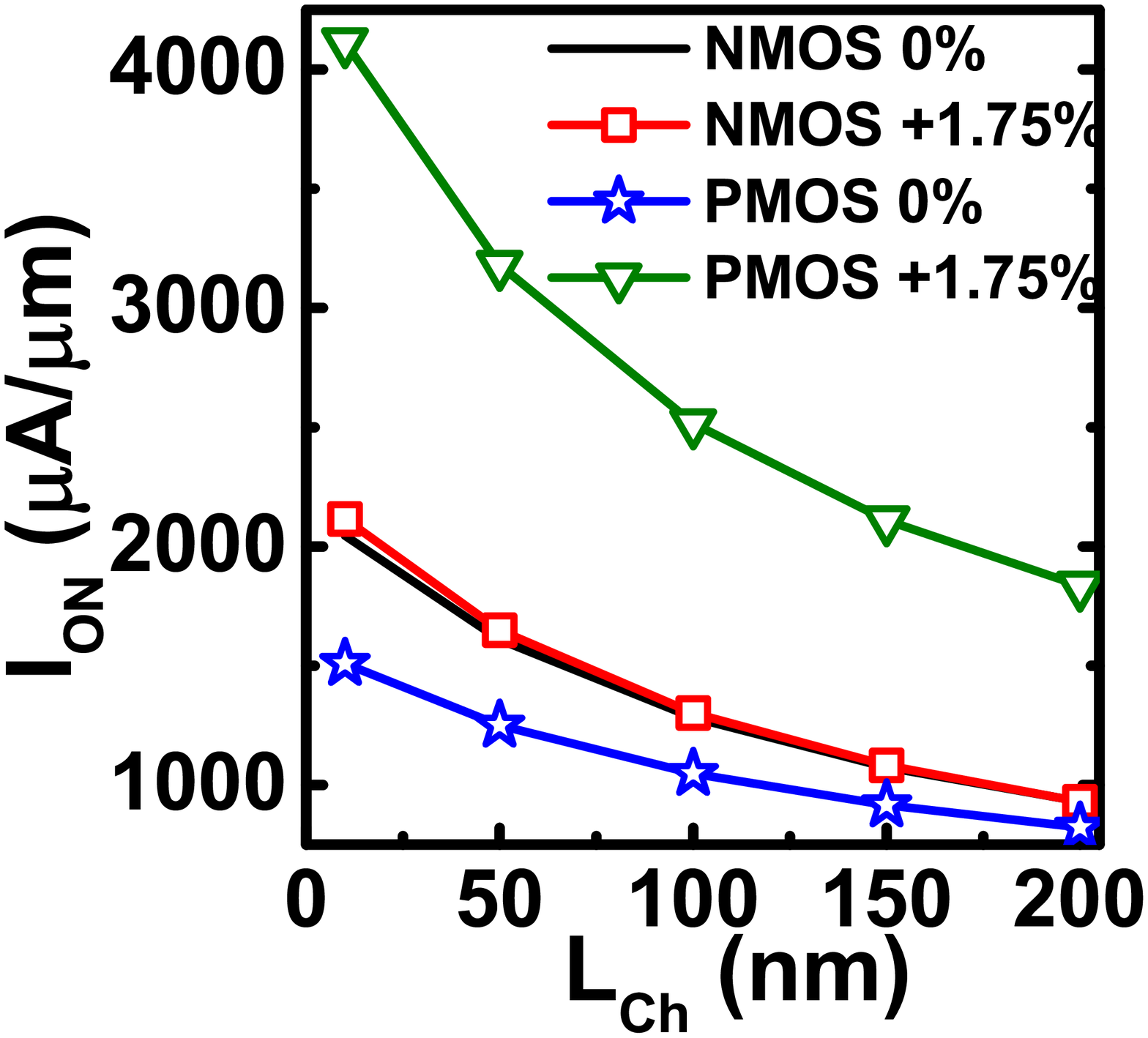}\label{a}}
\subfigure[]{\includegraphics[width=1.7in]{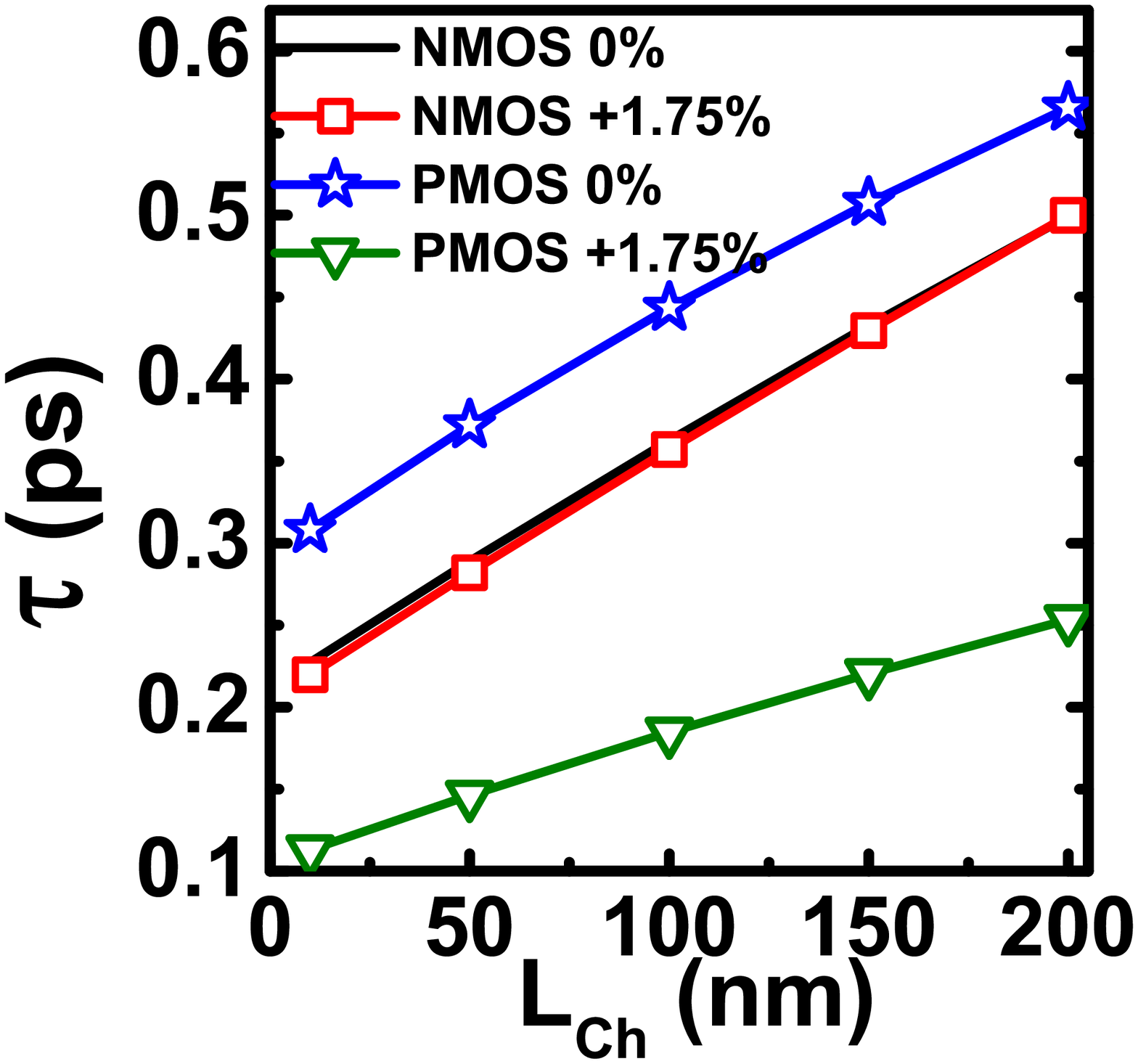}\label{b}}
\caption{Impact of channel length scaling on (a) the ON current and (b) the intrinsic delay time ($\tau$) of the MoS$_{2}$ NMOS and PMOS FETs.}
\end{center}
\label{fig_8}
\end{figure}
So far we have simulated all the results based on a 10nm channel length ($L_{Ch}$) MOSFET. In such short channel length MoS$_{2}$ FET, the carrier transport is purely ballistic in nature and there is no scattering involved in the channel. For our simulations we have considered such short $L_{Ch}$ in order to analyze the performance of this new alternate channel material, in the high performance technology node for next generation MOSFET application. However, most experimentally fabricated MoS$_{2}$ FET have $L_{Ch}$ in the range of several hundred nanometers to few microns\cite{Radisavlejvic,Kis}. In such long channel devices the carrier transport is no longer purely ballistic but quasi-ballistic in nature. In order to understand the performance of the strained MoS$_{2}$ FET in this region, we employ a projection method following Alam and Lake\cite{Alam} and Yoon et. al.\cite{Ganapathi}. Using equations (7)-(9) we simulate the projected currents in the long channel case. The ON current reduction and the increase in the intrinsic delay time for the NMOS and the PMOSFET in strained and relaxed conditions are shown in Fig. 8 (a) and 8 (b). As already discussed, we have considered uniaxial strain for the NMOS and biaxial strain for the PMOS device. For $L_{Ch}=200 nm$ the reduction in ON currents is 45$\%$ and 55$\%$ from the ballistic value, for the relaxed NMOS and the PMOS FET respectively. However for the strained PMOS, this reduction is 55$\%$ which is slightly higher than relaxed value. For the strained NMOS however, the $I_{ON}$ reduction remains the same. In fabricated long channel devices the reduction in drain current would be even higher owing to numerous defects and scattering centers formed during the processing. However, these projected currents give a good indication of the performance degradation for longer channel lengths.\\
As for the intrinsic delay time ($\tau$) is concerned it increases with an increasing $L_{Ch}$ for all the devices. The delay time increases by around 1.8 times for the PMOSFET and by 2.3 times for the NMOSFET as $L_{Ch}$ is increased to 200 nm.\\
\section{Conclusion}
The effect of varying tensile and compressive uniaxial and biaxial strain on the device performance of monolayer MoS$_{2}$ NMOS and PMOS FETs are studied herein. The material properties and the multi-band Hamiltonian of the channel, are evaluated using DFT. Using these parameters, the MOS device output have been simulated by self-consistent Poisson-Schr\"{o}dinger equations solution, under NEGF formalism. Our studies show uniaxial tensile strain to be beneficial for NMOSFET performance enhancement while biaxial tensile strain shows to significantly improve the PMOSFET performance. Compressive strain is found to be detrimental to performance of both NMOS and PMOS FET. We also observe that the PMOSFET performance enhancement is related to the transition of MoS$_{2}$ from direct band-gap to and indirect band-gap material under applied strain. By a projection method performance degradation of such strained MoS$_{2}$ FET in the quasi-ballistic region was also studied.
\section{Acknowledgement}
Dr. A. Sengupta thanks DST, Govt. of India, for the DST Post-doctoral Fellowship in Nano Science and Technology. R.K. Ghosh thanks UGC-CSIR, Govt. of India, for his Senior Research Fellowship.
%

\end{document}